\documentstyle[12pt]{article}
\begin{document}
\baselineskip 16pt plus 2pt minus 2pt
\newcommand{\beq}{\begin{equation}}
\newcommand{\eeq}{\end{equation}}
\newcommand{\beqa}{\begin{eqnarray}}
\newcommand{\eeqa}{\end{eqnarray}}
\newcommand{\dfrac}{\displaystyle \frac}
\renewcommand{\thefootnote}{\#\arabic{footnote}}
\newcommand{\ve}{\varepsilon}
\begin{titlepage}

\vspace{0.5cm}

\begin{center}

{\large  \bf WHAT IS A LOW--ENERGY THEOREM ? \footnote{Work supported in part
by FWF (Austria), Project No. P09505--PHY and
by HCM, EEC--Contract No. CHRX--CT920026 (EURODA$\Phi$NE)}}

\vspace{1.2cm}

{\large G. Ecker$^{\S ,\ddagger}$ and Ulf-G. Mei\ss ner$^{\dag
,\star,}$}\footnote{Address after October 1$^{st}$, 1994~: Universit\"at Bonn,
Institut f\"ur Theoretische Kernphysik, Nussallee 14-16, D--53115 Bonn,
Germany}

\vspace{0.7cm}

$^{\S}$Institut f\"ur Theoretische Physik, Universit\"{a}t Wien,
Boltzmanngasse 5, A-1090 Wien, Austria

\vspace{0.5cm}

$^{\dag}$Physique Th\'{e}orique, Centre de Recherches Nucl\'{e}aires et
Universit\'{e}
Louis Pasteur de Strasbourg, B.P. 28, F-67037 Strasbourg Cedex 2, France

\vspace{0.5cm}

email: $^\ddagger$ecker@ariel.pap.univie.ac.at, $\, $
$^\star$meissner@crnvax.in2p3.fr

\end{center}

\vspace{3cm}

\begin{center}

{\bf ABSTRACT}

\end{center}

\vspace{0.1cm}

\noindent
We discuss the meaning of low--energy theorems (LETs) in the
framework of the effective field theory of the standard model. Particular
emphasis is put on the LET for neutral pion photoproduction off nucleons at
threshold. The seemingly controversial situation surrounding this LET is
clarified.

\vspace{5cm}

\noindent CRN-94/52 \\
\noindent UWThPh-1994-33 \hfill September 1994\\
\noindent hep-ph/9409442

\vfill

\end{titlepage}
\section{Introduction}
\label{sec:intro}
Over the last few years, there has been some debate about the
low--energy theorem (LET) for the electric dipole
amplitude $E_{0+}$ in the reaction $\gamma p \to \pi^0 p$ at threshold. This
was spurred by the experimental findings that the LET derived in the early
1970's \cite{VZ} \cite{deB} seemed to be violated \cite{sac} \cite{beck}
leading to numerous re--examinations of the data as well as theoretical
reconsiderations of the LET. A lucid discussion of the status as of 1991 can
be found in the comment by Bernstein and Holstein \cite{bh}.
In this comment, we wish to
elaborate on certain aspects of LETs in the framework of the standard
model (SM). In particular, we propose an answer to the question~:
``What is a LET ?".

Let us first consider a well--known example of a LET
about which there is no discussion. Consider the
scattering of very soft photons on the proton, i.e., the Compton scattering
process $\gamma (k_1) + p(p_1) \to \gamma (k_2) + p(p_2)$
and denote by $\ve \, (\ve ')$
the polarization vector of the incoming (outgoing) photon. The transition
matrix element $T$ (normalized to $d\sigma / d\Omega = |T|^2$) can be
expanded in a Taylor series in the small parameter $\delta = |\vec{k_1}|/m$,
with $m$ the nucleon mass. In the forward  direction and in a gauge where
the polarization vectors have only space components, $T$ takes the form
\begin{equation}
T = c_0 \, \vec{\ve}\, ' \cdot \vec{\ve} + i \, c_1 \, \delta
\, \vec{\sigma} \cdot (
\vec{\ve}\, ' \times \vec{\ve} \, ) + {\cal O}(\delta^2) ~ .
\label{Comp}
\end{equation}
The parameter $\delta$ can be made arbitrarily small in the laboratory so
that the first two terms in the Taylor expansion (\ref{Comp}) dominate.
To be precise, the first one proportional to $c_0$ gives the low--energy
limit for the spin--averaged Compton amplitude, while the second ($\sim c_1$)
is of pure spin--flip type and can directly be detected in polarized
photon proton scattering. The pertinent LETs fix the values of $c_0$ and
$c_1$ in terms of measurable quantities \cite{low},
\begin{equation}
c_0 = - \frac{Z^2e^2}{4 \pi m} \, , \quad c_1 =
- \frac{Z^2e^2 \kappa_p^2}{8 \pi m}
\label{c01}
\end{equation}
with $Z =1$ the  charge of the proton and $\kappa_p = 1.793$ its anomalous
magnetic moment. To arrive at Eq.~(\ref{c01}), one only makes use of gauge
invariance and the fact that the $T$--matrix can be written in terms of a
time--ordered product of two conserved vector currents sandwiched between
proton states. The derivation proceeds by showing that for small
enough photon energies the matrix element is determined by the electromagnetic
form factor of the proton at $q^2 = 0$ \cite{low}.

Similar methods can be applied to other than the electromagnetic
currents. In strong interaction physics, a special role is played by the
axial--vector currents. The associated symmetries are spontaneously
broken giving rise to the Goldstone matrix elements
\beq
\langle 0|A^a_\mu(0)|\pi^b(p)\rangle =
i \delta_{ab} F_\pi p_\mu
\label{Gme}
\eeq
where $a,b$ are isospin indices and $F_\pi \simeq 93$ MeV  is the
pion decay constant. In the chiral limit (vanishing quark masses)
the massless pions play a similar role as the photon and many
LETs have been derived for ``soft pions".
In light of the previous discussion on Compton
scattering, the most obvious one is Weinberg's prediction for elastic
$\pi p$ scattering \cite{wein68}. We only need the following translations~:
\begin{equation}
<p| T \, j_\mu^{\rm em} (x) j_\nu^{\rm em} (0)|p> \, \, \to \, \,
<p| T \, A_\mu^{\pi^+} (x) A_\nu^{\pi^-} (0)|p> ~,
\end{equation}
\begin{equation}
\partial^\mu j_\mu^{\rm em} = 0 \, \, \to \, \,
\partial^\mu A_\mu^{\pi^-}  = 0  ~.
\end{equation}

In contrast to photons, pions are not massless in the real
world. It is therefore interesting to find out how the LETs for
soft pions are modified in the presence of non--zero pion masses
(due to non--vanishing quark masses). In the old days of current
algebra, a lot of emphasis was put on the PCAC
(Partial Conservation of the Axial--Vector Current) relation,
consistent with the Goldstone matrix element (\ref{Gme}),
\begin{equation}
\partial^\mu A^a_\mu = M_\pi^2 F_\pi \phi^a_\pi ~ ,
\label{PCAC}
\end{equation}
where $\phi^a_\pi$ denotes the pion field and $M_\pi \simeq 140 $ MeV
is the pion mass. Although the precise meaning of (\ref{PCAC}) has
long been understood \cite{col}, it does not offer a systematic
method to calculate higher orders in the momentum and mass expansion
of LETs. The derivation of non--leading terms in the days of
current algebra and PCAC was more an art than a science, often
involving dangerous procedures like off--shell
extrapolations of amplitudes (see also Sect.~\ref{sec:photo}).

The modern developments in this field have replaced the old notions
by the effective field theory (EFT) of the SM incorporating all the
symmetries of the SM including the spontaneously broken chiral
symmetry. This framework to be sketched in Sect.~\ref{sec:CHPT} allows
for a systematic expansion of amplitudes
and Green functions in terms of momenta and meson masses. One
recovers all the old LETs that are rightfully called theorems, but
one does not reproduce some of the old results that were based on
unjustified assumptions not valid in the SM. After the general definition
of a LET in the new framework in Sect.~\ref{sec:CHPT}
emphasizing the concept of chiral power counting, we briefly treat $\pi\pi$
scattering as a special example in Sect.~\ref{sec:pipi}. In
Sect.~\ref{sec:comp} we reconsider the LET for Compton scattering in the
framework of heavy baryon chiral perturbation theory. Our main
concern, however, will be to clarify the status of the LETs for
$E_{0+}$ in $\gamma N \to \pi^0 N$ at threshold (see
also the discussion in Ref.~\cite{bklm}) in
Sect.~\ref{sec:photo} and to discuss some of
the pitfalls of the old methods that can be avoided with modern
techniques.

\section{Definition of low--energy theorems}
\label{sec:CHPT}
Chiral perturbation theory (CHPT) is the EFT of the SM at low
energies in the hadronic sector. Since as an EFT it contains all terms allowed
by the symmetries of the underlying theory \cite{wein79}, it should be viewed
as a direct consequence of the SM itself. The two main assumptions
underlying CHPT are that
\begin{enumerate}
\item[(i)] the masses of the light quarks $u$, $d$ (and possibly
$s$) can be
treated as perturbations (i.e., they are small compared to a typical
hadronic scale of 1 GeV) and  that
\item[(ii)] in the limit of zero quark masses, the chiral symmetry is
spontaneously broken  to its vectorial subgroup. The resulting Goldstone
bosons are the pseudoscalar mesons (pions, kaons and eta).
\end{enumerate}

CHPT is a systematic low--energy expansion around the
chiral limit \cite{wein79} \cite{GL1} \cite{GL2} \cite{leut}. It is a
well--defined quantum field theory although it has to be renormalized
order by order. Beyond leading order, one has to include loop diagrams to
 restore unitarity perturbatively. Furthermore, Green functions calculated in
CHPT at a given order contain certain parameters that are not constrained by
the symmetries, the so--called low--energy constants (LECs).
At each order in the chiral expansion, those LECs have to be
determined  from phenomenology (or can be estimated with some
model dependent assumptions).
For a review of the wide field of applications of CHPT, see, e.g., Ref.~\cite
{UGM}.

In the baryon sector, a
complication arises from the fact that the baryon mass $m$ does not vanish
in the chiral limit \cite{GSS}. Stated differently, only baryon three--momenta
 can be small compared to the hadronic scale. To restore the
correspondence between the loop and the energy expansion valid in
the meson sector, one can reformulate baryon CHPT \cite{GSS} in
analogy to heavy quark effective theory to shift the
troublesome mass term from the baryon propagator to a string of interaction
vertices with increasing powers of $1/m$ \cite{JM}. The
procedure is reminiscent of the well--known Foldy--Wouthuysen
transformation and is called heavy baryon chiral perturbation theory
(HBCHPT). The baryon four--momentum is written as $p_\mu = m v_\mu + l_\mu$,
with $v_\mu$ the four--velocity and $l_\mu$ a small off--shell momentum, $v
\cdot l \ll m$. The Dirac equation for the velocity--dependent
baryon field $B_v$ takes the form $i v \cdot \partial B_v = 0$ to lowest
order in $1/m$. This allows for a consistent chiral counting as described
below.

We are now ready to address the central question of this comment~:

\begin{center}
{\bf What is a LET ?} \\

\medskip

 {\bf L}(OW) {\bf E}(NERGY) {\bf T}(HEOREM) OF
 ${\cal O}(p^n)$   \\

 $\equiv$ GENERAL PREDICTION OF CHPT TO ${\cal O}(p^n)$  \\

\medskip

\end{center}
\noindent As will be explained below, $p$ stands for a small momentum or mass
characterizing the chiral expansion.  By general prediction we mean a
strict consequence of the SM depending on some LECs like
$F_\pi, m, g_A, \kappa_p, \ldots$, but without any model assumption for these
parameters. This definition contains a
precise prescription how to obtain higher--order corrections to
leading--order LETs and it should therefore be generally
adopted for hadronic processes at low energies. Although we have
formulated the procedure with the SM in mind, the obtained LETs
are actually more general. Since one only uses the symmetries of
the SM to derive general results of CHPT, those results hold in fact
in any theory that shares the symmetries of the SM. This general
aspect of a LET is less relevant today than 30 years
ago, but it should be kept in mind.

We have to be a little more precise what is meant by a result of
${\cal O}(p^n)$. From the outset, one can distinguish between
an expansion in momenta (CHPT is a low--energy effective theory)
and an expansion around the chiral limit in terms of quark masses.
These two expansions become related by expressing the pseudoscalar
meson masses in terms of the quark masses. We adopt here the standard
assumption supported by the success of the Gell-Mann--Okubo mass
formula for the pseudoscalar octet that the dominant
contributions to the squares of the meson masses are linear in the
quark masses, e.g.,
\beq
M^2_{\pi^+} = B (m_u + m_d)[1+{\cal O}(m_{quark})]~.
\label{masses}
\eeq
The constant $B$ is related to the quark condensate and is
assumed to be non--vanishing in the chiral limit (supported by lattice data).
In this case, Eq.~(\ref{masses}) implies the standard chiral
counting where quark masses count as ${\cal O}(p^2)$. If one declares
the Gell-Mann--Okubo formula to be a numerical accident, one can
envisage a situation where $B$ is very small or even zero so
that the higher--order terms in (\ref{masses}) could be
dominant. The proponents of ``Generalized CHPT" \cite{Stern}
account for this possibility by considering the quark masses as objects
of ${\cal O}(p)$. In practice, this means that at any given
order the CHPT generalizers include some additional terms which
would only appear in higher orders in the standard counting.
Since there is at this time no phenomenological necessity
to include those terms (with their associated unknown LECs), we
stick to the standard procedure. Of course, a difference can
only appear in LETs where symmetry breaking terms in the chiral Lagrangian
contribute. Anticipating the examples discussed below, the generalized
counting affects $\pi\pi$ scattering already at ${\cal O}(p^2)$
(Sect.~\ref{sec:pipi}), but it does not modify the LETs for Compton
scattering (Sect.~\ref{sec:comp}) or for neutral pion
photoproduction (Sect.~\ref{sec:photo}).

The soft--photon theorems, e.g., for Compton scattering \cite{low},
involve the limit of small photon momenta, with all other momenta
remaining fixed. Therefore, they hold to all orders in the non--photonic
momenta and masses. In the low--energy expansion of CHPT, on the
other hand, the ratios of all small momenta and pseudoscalar meson
masses are held fixed. Of course, the soft--photon theorems are also
valid in CHPT as in any gauge invariant quantum field theory. We shall
come back to this difference of low--energy limits in Sect.~\ref{sec:comp}
in the derivation of the LET for Compton scattering.

To calculate a LET to a given order, it is useful to have
a compact expression for the chiral power counting \cite{wein79} \cite{ecker}.
We restrict ourselves to purely mesonic or single--nucleon processes. Any
amplitude for a given physical process has a certain {\bf chiral} {\bf
dimension} $D$ which keeps track of the powers of external momenta and meson
masses. The building blocks to calculate this chiral dimension
from a general Feynman diagram in the CHPT loop expansion are
\begin{enumerate}
\item[(i)] $I_M$ meson propagators $\sim 1/(k^2 -M^2)$ (with $M$ the meson
mass) of dimension $D= -2$,
\item[(ii)] $I_B$ baryon propagators $\sim 1/ v \cdot k$ (in HBCHPT) with
$D= -1$,
\item[(iii)] $N_d^M$ mesonic vertices with $d =2,4,6, \ldots$ and
\item[(iv)] $N_d^{MB}$ meson--baryon vertices with $d = 1,2,3, \ldots$.
\end{enumerate}
Putting these together, the chiral dimension $D$ of a given amplitude reads
\begin{equation}
D =4L - 2I_M - I_B + \sum_d d( \,  N_d^M + N_d^{MB} \, )
\end{equation}
with $L$ the number of loops. For connected diagrams, we can use
the general topological relation
\begin{equation}
L = I_M + I_B -  \sum_d ( \,  N_d^M + N_d^{MB} \, ) + 1
\end{equation}
to eliminate $I_M$~:
\begin{equation}
D =2L + 2 + I_B + \sum_d (d-2)  N_d^M + \sum_d (d-2) N_d^{MB} ~ .
\label{DLgen}
\end{equation}
Lorentz invariance and chiral symmetry demand $d \ge 2$ for mesonic
interactions and thus the
term $\sum_d (d-2)  N_d^M$ is non--negative. Therefore, in the absence
of baryon fields, Eq.~(\ref{DLgen}) simplifies to \cite{wein79}
\begin{equation}
D =2L + 2 + \sum_d (d-2)  N_d^M  \, \ge 2L + 2 ~ .
\label{DLmeson}
\end{equation}
To lowest order $p^2$, one has to deal with tree diagrams
($L=0$) only. Loops are suppressed by powers of $p^{2L}$.

The other case of interest for us has a single baryon line running through
the diagram (i.e., there is exactly one baryon in the in-- and one baryon
in the out--state). In this case, the identity
\begin{equation}
\sum_d N_d^{MB} = I_B + 1
\end{equation}
holds leading to \cite{ecker}
\begin{equation}
D =2L + 1  + \sum_d (d-2)  N_d^M + \sum_d (d-1)  N_d^{MB}  \, \ge 2L + 1 ~ .
\label{DLMB}
\end{equation}
Therefore, tree diagrams start to contribute at order $p$ and one--loop
graphs at order $p^3$. Obviously, the relations involving
baryons are only valid in HBCHPT.

Let us now consider diagrams with $N_\gamma$ external photons.\footnote
{We remind the reader that CHPT as discussed here has {\bf only} external
photons.}
Since gauge fields like the electromagnetic field appear
in covariant derivatives, their chiral dimension is obviously $D=1$.
We therefore write the chiral dimension of a general amplitude
with $N_\gamma$ photons as
\begin{equation}
D = D_L + N_\gamma  ~,
\end{equation}
where $D_L$ is the degree of homogeneity of the (Feynman) amplitude $A$ as
a function of external momenta ($p$) and meson masses ($M$) in the
following sense \footnote{A
similar observation has been made by Rho \cite{rho} in the context
of meson exchange currents.}:
\begin{equation}
A(p,M;C_i^r(\mu),\mu/M) = M^{D_L} \, A ( p/M , 1;C_i^r(\mu), \mu/M )  ~ ,
\end{equation}
where $\mu$ is an arbitrary renormalization scale and $C_i^r(\mu)$
denote renormalized LECs. From now on, we suppress the explicit dependence
on the renormalization scale and on the LECs. Since the total amplitude
is independent of the arbitrary scale $\mu$, one may in particular
choose $\mu=M$.
Note that $A(p,M)$ has also a certain physical dimension (which is
of course independent of the number of loops and is therefore in
general different from $D_L$). The correct
physical dimension is ensured by appropriate factors of $F_\pi$ and $m$
in the denominators as will become evident from the following examples.

In the remaining sections, we always consider chiral $SU(2)$ in
the isospin limit ($m_u=m_d$).

\section{Pion--pion scattering}
\label{sec:pipi}

We first consider the mesonic sector. The purest reaction to test chiral
dynamics is elastic $\pi \pi$ scattering in the threshold region. It involves
exclusively Goldstone bosons and the expansion parameters $E^2 / (4\pi
F_\pi)^2$  and $M_\pi^2 / (4\pi F_\pi)^2 \simeq
0.014$ are small. The $\pi \pi$ scattering amplitude can be written in terms of
a single invariant function, conventionally called $A(s,t,u)$ where $s,t$ and
$u$ are the Mandelstam variables. The chiral expansion of $A(s,t,u)$ takes the
form
$$A(s,t,u) = A^{(2)}(s,t,u) + A^{(4)}(s,t,u) + {\cal O}(p^6)$$
where $p^6$ denotes terms of the type $E^6, E^4 M_\pi^2, E^2 M_\pi^4$ or
$M_\pi^6$ ($E^2 = s,t~ {\rm or}~u$). Since no external photons are involved,
we have $D = D_L$. From Eq.~(\ref{DLmeson}) we read
off that to lowest order $L=0$ and $d=2$ only (tree diagrams with insertions
from the lowest--order chiral Lagrangian ${\cal L}_2$). The corresponding LET
of order $p^2$ was derived by Weinberg \cite{wein66}~:
\begin{equation}
A^{(2)}(s,t,u) = \frac{s - M_\pi^2}{F_\pi^2} = \frac{M_\pi^2}{F_\pi^2} \left(
\frac{s}{M_\pi^2} -1 \right)~.
\label{A2}
\end{equation}
We notice that the degree of homogeneity $D_L =2$ indeed differs from the
physical dimension of the amplitude which in this case is dimensionless.
If one projects the amplitude (\ref{A2}) onto
channels with isopin $I$ and angular momentum $l$ of the two--pion
system and expands the corresponding partial waves $t_l^I (s)$ in powers of
the pion three--momentum, one can read off the well--known LET for the $I=0$
$S$--wave scattering length \cite{wein66}~:
\begin{equation}
a_0^0 = \frac{7 M_\pi^2}{32 \pi F_\pi^2} = 0.16 ~,
\label{e20}
\end{equation}
to be compared with the empirical value of $a^0_{0,{\rm exp}} = 0.26 \pm 0.05$
\cite{pet}.

At next--to--leading order, $D=D_L=4$, one has two types of contributions.
First, there are (divergent) loop diagrams with $L =1$ and $N_d^M = 0$
for $d > 2$ and second, counterterms with $L =0, N_4^M =1$ and
$N_d^M = 0 ~(d > 4)$. The
latter involve some LECs denoted $\bar{l}_i$ in the $SU(2)$ analysis
\cite{GL1}. The complete
amplitude of ${\cal O}(p^4)$ is of the form
\begin{equation}
A^{(4)}(s,t,u) = \frac{M_\pi^4}{F_\pi^4} \widehat{A} \left(
\frac{s}{M_\pi^2}, \frac{t}{M_\pi^2}, \frac{u}{M_\pi^2} \right)~.
\label{A4}
\end{equation}
The LET of
order $p^4$ for the $\pi \pi$ scattering amplitude is simply the sum of
eqs.~(\ref{A2}) and (\ref{A4}). In particular, the LET for $a_0^0$ to order
$p^4$ reads \cite{GL1}
\begin{equation}
a_0^0 = \frac{7 M_\pi^2}{32 \pi F_\pi^2} \biggl\lbrace 1 +
\frac{M_\pi^2}{3} <r^2>_S^\pi - \frac{M_\pi^2}{672\pi^2 F_\pi^2} ( 15
\bar{\ell}_3 -353) \biggr\rbrace +\frac{25}{4}M_\pi^4 (a_2^0 + 2a_2^2)
 = 0.20 \pm 0.01
\label{a00}
\end{equation}
with $<r^2>_S^\pi ~\simeq 0.6$ fm$^2$ the scalar radius of the pion
\cite{DGL90} and
$a_2^0, a_2^2$ the $D$--wave scattering lengths. An estimate for the LEC
$\bar{\ell}_3$ can be found in Ref.~\cite{GL1}.
The precise prediction (\ref{a00})
awaits an equally accurate empirical determination (for a more detailed
discussion on the relevance of pinning down $a_0^0$, see, e.g.,
Ref.~\cite{UGM}).

\section{Compton scattering revisited}
\label{sec:comp}

In this section, we rederive and extend the LET for spin--averaged nucleon
Compton scattering in the framework of HBCHPT \cite{BKKM}. Consider the
spin--averaged Compton amplitude in forward direction (in the Coulomb
gauge $\ve \cdot v = 0$)
\beq
e^2 \ve^\mu \ve^\nu
\frac{1}{4} {\rm Tr} \biggl[ (1 + \gamma_\lambda v^\lambda) T_{\mu \nu} (v,k)
\biggr] = e^2 \biggl[ \ve^2 U(\omega ) + (\ve \cdot k )^2
V(\omega ) \biggr]
\eeq
with $\omega = v \cdot k$ ($k$ is the photon momentum) and
\beq
T_{\mu \nu} (v,k) = \int d^4 k \, {\rm e}^{ik \cdot x} \, < N(v)| T j^{\rm
em}_\mu (x) j_\nu^{\rm em} (0) |N(v)>~.
\eeq
All dynamical information is
contained in the functions $U(\omega)$ and $V (\omega )$. We
only consider $U(\omega)$ here and refer to Ref.~\cite{BKKM} for
the calculation of both $U(\omega)$ and $V(\omega)$. In the
Thomson limit, only $U(0)$ contributes to the amplitude.

In the forward direction, the only quantities with non--zero chiral
dimension are $\omega$ and $M_\pi$. In order to make this dependence
explicit, we write $U(\omega,M_\pi)$ instead of $U(\omega)$. With
$N_\gamma = 2$ external photons, the degree of homogeneity $D_L$
for a given CHPT contribution to $U(\omega,M_\pi)$ follows from
Eq.~(\ref{DLMB})~:
\beq
D_L =2L - 1  + \sum_d (d-2)  N_d^M + \sum_d (d-1)  N_d^{MB}  \, \ge - 1 ~ .
\label{DLC}
\eeq
Therefore, the chiral expansion of $U(\omega,M_\pi)$ takes the following
general form~:
\beq
U(\omega,M_\pi) = \sum_{D_L\ge -1} \omega^{D_L} f_{D_L}(\omega/M_\pi)~.
\label{Uce}
\eeq

The following arguments illuminate the difference and the
interplay between the soft--photon limit and the low--energy expansion
of CHPT. Let us consider first the leading terms in the chiral expansion
(\ref{Uce})~:
\beq
U(\omega,M_\pi) = {1\over \omega} f_{-1}(\omega/M_\pi) +
 f_0(\omega/M_\pi) + {\cal O}(p^3)~.
\eeq
Eq.~(\ref{DLC}) tells us that only tree diagrams can contribute
to the first two terms. However, the relevant tree diagrams
shown in Fig.~1 do not contain pion lines. Consequently, the functions
$f_{-1}$, $f_0$ cannot depend on $M_\pi$ and are therefore constants.
Since the soft--photon theorem \cite{low} requires $U(0,M_\pi)$
to be finite, $f_{-1}$ must actually vanish and the chiral
expansion of $U(\omega,M_\pi)$ can be written as
\beq
U(\omega,M_\pi) = f_0 + \sum_{D_L\ge 1} \omega^{D_L} f_{D_L}(\omega/M_\pi)~.
\label{Uce2}
\eeq
But the soft--photon theorem yields additional information~:
since the Compton amplitude is independent of $M_\pi$ in the Thomson
limit and since there is no term linear in $\omega$ in the spin--averaged
amplitude, we find
\beq
\lim_{\omega \to 0}~ \omega^{n-1} f_n(\omega/M_\pi) = 0 \qquad
(n\ge 1) \label{spl}
\eeq
implying in particular that the constant $f_0$ describes the
Thomson limit~:
\beq
U(0,M_\pi) = f_0~.
\eeq

Let us now verify these results by explicit calculation.
In the Coulomb gauge, there is no direct
photon--nucleon coupling from the lowest--order effective Lagrangian
${\cal L}_{\pi N}^{(1)}$ since it is proportional to $\ve \cdot v$.
Consequently, the Born diagrams a,b in Fig.~1 vanish so that indeed
$f_{-1}=0$. On the other hand, the expansion of the relativistic Dirac
Lagrangian leads to terms of the type $D^2 /2m$ and $(v \cdot D)^2 /2m$
where $D_\mu$ is a covariant derivative.
Notice that although these terms belong to ${\cal L}_{\pi N}^{(2)}$, they
do not contain novel LECs since they are of purely kinematical origin.
These terms lead to a Feynman insertion (Fig.~1c) of the form
\beq
i \frac{e^2}{m} \frac{1}{2} (1 + \tau_3 ) \biggl[ \ve^2 -( \ve
\cdot v)^2 \biggr] = i \frac{e^2 Z^2}{m} \,  \ve^2
\eeq
producing the desired result $f_0 = Z^2 / m$, the Thomson limit.

At the next order in the chiral expansion, ${\cal O}(p^3)$ ($D_L = 1$),
the function $f_1(\omega/M_\pi)$ is given by the
finite sum of 9 one--loop diagrams \cite{BKM} \cite{BKKM}. According
to Eq.~(\ref{spl}), $f_1$ vanishes for $\omega \to 0$. The term linear
in $\omega/M_\pi$ yields the leading contribution to the sum of the
electric and magnetic polarizabilities of the nucleon, defined by
the second--order Taylor coefficient in the expansion of $U(\omega,M_\pi)$
in $\omega$~:
\beq
f_1(\omega/M_\pi) = - {11 g_A^2 \omega\over 192 \pi F_\pi^2 M_\pi}
+ {\cal O}(\omega^2)~,
\eeq
where $g_A$ is the nucleon axial--vector coupling constant.
The $1/M_\pi$ behaviour should not come as a surprise
-- in the chiral limit the  pion cloud becomes long--ranged (instead of being
Yukawa--suppressed) so that the polarizabilities explode.
This behaviour is specific to the leading contribution
of ${\cal O}(p^3)$. In fact, from the general form (\ref{Uce2}) one
immediately derives that the contribution of ${\cal O}(p^n)$
($D_L = n - 2$) to the polarizabilities is of the form $c_n M_\pi^{n-4}$
($n\ge 3$), where $c_n$ is a constant that may be zero.

One can perform a similar analysis for the amplitude
$V(\omega)$ and for the spin--flip amplitude. We do not
discuss these amplitudes here but refer the reader to Ref.~\cite{BKKM}
for details.

\section{Neutral pion photoproduction at threshold}
\label{sec:photo}

We consider the processes
$$\gamma N \to \pi^0 N \qquad (N=p,n)$$
at threshold, i.e., for vanishing three--momentum of the pion
in the nucleon rest frame. At
threshold, only the electric dipole amplitude $E_{0+}$ survives and
the only quantity with non--zero chiral dimension is $M_\pi$.
In the usual conventions,
$E_{0+}$ has physical dimension $-1$ and it can therefore be
written as
\beq
E_{0+} = {e g_A \over F}~A\left( {M_\pi\over m},{M_\pi\over F}\right)~,
\eeq
where $F$ is the pion decay constant in the chiral limit.
The dimensionless amplitude $A$ will be expressed as a power
series in $M_\pi$. The various parts are characterized by the
degree of homogeneity (in $M_\pi$) $D_L$ according to the chiral
expansion. Since $N_\gamma=1$ in the present case, we obtain from
Eq.~(\ref{DLMB})
\beq
D_L = D-1 = 2L + \sum_d (d-2)  N_d^M + \sum_d (d-1)  N_d^{MB} ~ .
\eeq
For the LET of ${\cal O}(p^3)$ in question, only lowest--order mesonic vertices
($d=2$) will appear. Therefore, in this case the general formula for $D_L$
takes the simpler form
\beq
D_L = 2L + \sum_d (d-1)  N_d^{MB} ~ .
\label{DLphoto}
\eeq

We now discuss the chiral expansion of $E_{0+}$ step by step, referring
to the literature \cite{BGKM} \cite{BKM1} \cite{BKKM} for the actual
calculation.

\begin{center}
${\bf D_L = 0}$
\end{center}
\noindent From Eq.~(\ref{DLphoto}) we conclude that only tree diagrams ($L=0$)
with vertices from the ${\cal O}(p)$ chiral pion--nucleon
Lagrangian ${\cal L}_{\pi N}^{(1)}$ can contribute. At
threshold, the only diagram is the Kroll--Ruderman contact term
\cite{KR} where both the pion and the photon emanate from the same
vertex. However, this vertex only exists for charged pions.
Thus, there is no term with $D_L=0$ for neutral pion production.

\begin{center}
${\bf D_L = 1}$
\end{center}
\noindent In HBCHPT, the relevant tree diagram (remember that $L=0$ for
$D_L < 2$) looks exactly like the Kroll--Ruderman diagram,
except that now the vertex comes from the ${\cal O}(p^2)$ pion--nucleon
Lagrangian. In the relativistic formulation \cite{GSS}, the contribution
is due to the normal scattering (and crossed) diagrams retaining
only the nucleon mass in the nucleon propagator. HBCHPT replaces
these diagrams by a contact term proportional to $1/m$. From the
relativistic description it is clear that this contribution is proportional
to the nucleon charge. For the neutron, both the $D_L=0$ and the
$D_L=1$ pieces vanish.

\begin{center}
${\bf D_L = 2}$
\end{center}
\noindent The master formula (\ref{DLphoto}) allows in principle for three
types of contributions~:
\begin{enumerate}
\item[(a)] Tree level ($L=0$) diagrams with a single vertex
of ${\cal O}(p^3)$ ($N_3^{MB}=1$,  but all other $N_d^{MB}=0$ for
$d > 1$). Although such vertices exist, they can be shown not to
contribute to neutral pion photoproduction at threshold \cite{BKKM}.
This has another important implication~: the loop contribution
to be discussed below must be finite because chiral and gauge invariance
do not permit appropriate counterterms of ${\cal O}(p^3)$.
\item[(b)] There are non--vanishing Born diagrams with a nucleon
propagator between ${\cal O}(p^2)$ $\gamma NN$ and $\pi NN$ couplings,
respectively ($L=0$, $N_2^{MB}=2$, $N_d^{MB}=0 ~(d>2))$ . The
$\gamma NN$ coupling is proportional to a LEC of HBCHPT, the
magnetic moment of the nucleon (in the chiral limit).
\item[(c)] Finally, and this is the piece that has generated a considerable
amount of paper and some heated discussions, there is a one--loop
contribution ($L=1$) with leading--order vertices only
($N_d^{MB}=0 ~(d>1)$).  It is considerably easier to work out the
relevant diagrams in HBCHPT \cite{BKKM} than in the original derivation
\cite{BGKM} \cite{BKM1}. In fact, at threshold only the so--called triangle
diagrams shown in Fig.~2 survive out
of some 60 diagrams. The main reason for the enormous simplification
in HBCHPT is that one can choose a gauge without a direct $\gamma NN$
coupling of lowest order and that there is no direct coupling of
the produced $\pi^0$ to the nucleon at threshold. As already noted,
the loop contributions are finite and they are identical for proton
and neutron. They were omitted in the original version of the LET
\cite{VZ} \cite{deB} and in many later rederivations.
\end{enumerate}

The full LETs of ${\cal O}(p^3)$ are given by \cite{BGKM}

\renewcommand{\arraystretch}{3}
$$
\begin{tabular}{|crcccccccr|} \hline
$E_{0+}(\pi^0 p)=  $ &  $- \dfrac{e g_A}{8\pi F_\pi}\biggl[ \biggr.$ & 0 & + &
$\dfrac{M_\pi}{m}$ & -- & $\dfrac{M_\pi^2}{2 m^2}~(3+\kappa_p)$ & -- &
$\dfrac{M_\pi^2}{16 F_\pi^2}$ & + $\biggl.{\cal O}(M_\pi^3)\biggr]$ \\[0pt]
$D_L ~:$ & & 0 & & 1 & & $2_b$ & & $2_c$ & \\[0pt]
$E_{0+}(\pi^0 n)=  $ &  $- \dfrac{e g_A}{8\pi F_\pi}\biggl[ \biggr.$ & 0 & + &
0 & + & $\dfrac{M_\pi^2}{2 m^2}~\kappa_n$ & -- &
$\dfrac{M_\pi^2}{16 F_\pi^2}$ & + $\biggl. {\cal O}(M_\pi^3)\biggr]$
\\[10pt]
\hline
\end{tabular}
$$
\renewcommand{\arraystretch}{1}

\noindent Two comments are in order here~:
\begin{enumerate}
\item[(i)] There is a kinematical factor in the relation between
the electric dipole amplitude and the Feynman amplitude depending
on $M_\pi$ . Expanding this factor in $M_\pi/m$ affects the
${\cal O}(M_\pi^2)$ term in the case of the proton. This explains the
factor $3+\kappa_p$ instead of
$1+\kappa_p$. For the neutron, this does not influence the LET
to ${\cal O}(M_\pi^2)$ because there is no term with $D_L=1$.
\item[(ii)] All LECs appearing in the LETs are the physical quantities
$g_A$, $m$, $F_\pi$, $\kappa_p$, $\kappa_n$
although the effective chiral Lagrangian contains the corresponding
quantities in the chiral limit. It is a major conceptual
advantage of HBCHPT that the relation between the physical and
the chiral limit values of all these parameters is such that the
differences can only appear in the higher--order terms denoted as
${\cal O}(M_\pi^3)$ (see also below). To prove the analogous statement
in the relativistic
formulation is much more cumbersome. In fact, most of the loop
contributions encountered in the relativistic approach renormalize
the various constants to their physical values \cite{BGKM} \cite{BKM1}.
Of course, the final result is the same in both approaches.
\end{enumerate}

The derivation of LETs sketched above is based on a well--defined
quantum field theory where each step can be checked explicitly. Nevertheless,
the corrected LETs have been questioned by several authors. We
find it instructive to discuss some of the arguments and assumptions
that have been used to derive or rederive the original LETs. Generically,
those derivations are based on more or less plausible assumptions that
qualify the results as LEGs (low--energy guesses) rather than
LETs. Since we have CHPT at our disposal as the effective low--energy
representation of the SM, we can actually check whether or not those
assumptions hold in the SM. The following list is not
meant to be exhaustive nor is it intended to be a compilation
of mistakes in the published literature.
It should rather be viewed as a collection of pitfalls that
should be looked out for when extending LETs beyond leading order.

\subsection*{(a) Analyticity assumption}
The original derivations \cite{VZ} \cite{deB} and some later rederivations
\cite{Naus1} \cite{Naus2} \cite{Kamal} \cite{bh} used a Taylor expansion
of amplitudes in the variables $\nu, \nu_B$ (linear combinations of the
usual Mandelstam variables $s$ and $u$). The seemingly plausible assumption
that the coefficients of this expansion are analytic in $M_\pi$ leads
directly to the original LEG. In fact, in Ref.~\cite{VZ} it was explicitly
spelled out that this is a necessary assumption for the LEG
to hold. However, as shown in
Ref.~\cite{BGKM}, this assumption does not hold in QCD. Due to the Goldstone
nature of the pion, some Taylor coefficients diverge in the chiral
limit. This happens precisely in the loop contributions ($D_L=2$)
which generate infrared divergences in some coefficients.
The threshold amplitude itself is perfectly well--behaved in the chiral limit.

\subsection*{(b) External versus internal pion mass}
It has been suggested \cite{Naus2} \cite{Naus3} \cite{Scherer}
that there is a basic difference between the external,
kinematical pion mass $M_\pi$ and the internal mass $\bar M_\pi$
appearing in the pion propagators in loop diagrams. The assumption
is that $\bar M_\pi$ appears only in relations between unrenormalized
and renormalized quantities. Therefore, expressing everything in
measurable, renormalized quantities, no trace of $\bar M_\pi$ is
left and one recovers the original LEG since the loop
contribution is to be dropped by assumption.

Let us investigate this assumption in detail within HBCHPT. Denoting
unrenormalized quantities (the parameters in
the effective chiral Lagrangian) with a superscript $\circ$, one
finds the following relations between physical, renormalized quantities
and their unrenormalized counterparts~:
\beq
Q = ~\stackrel{\circ}{Q} [1+{\cal O}(m_q)] \equiv
 ~\stackrel{\circ}{Q} [1+{\cal O}(p^2)] \, ,
\quad Q = M_\pi, F_\pi, m, g_A
\eeq
\beq
\kappa = ~\stackrel{\circ}{\kappa} [1+{\cal O}(m_q^{1/2})] \equiv
 ~\stackrel{\circ}{\kappa} [1+{\cal O}(p)]~.
\eeq
As already emphasized before, renormalization in
the framework of HBCHPT can therefore only affect terms of
${\cal O}(M_\pi^3)$ in
the LETs for $\pi^0$ photoproduction at threshold. Thus, the loop
contribution to the LETs cannot be a renormalization effect.

There is a more fundamental objection to the distinction between external
and internal pion masses \cite{Heiri}. QCD does not offer a consistent
procedure for $M_\pi \to 0$ with $\bar M_\pi$ remaining finite. The
only tunable mass parameters in QCD are the quark masses.
Letting the quark masses tend to zero makes all pion masses vanish,
whether they be external or internal.

\subsection*{(c) Off--shell expansion}
The inadmissible distinction between external and internal pion masses
can also appear in an off--shell extrapolation of the amplitude.
Davidson has contrasted the expansion in $M_\pi$ with
a so--called $\omega$ expansion \cite{David}. Keeping $M_\pi$ fixed, he
sets the three--momentum $\vec p_\pi =0$ and expands in the pion
energy $E_\pi = \omega$. Obviously, for $\omega \ne M_\pi$ this
implies an off--shell extrapolation of the scattering amplitude.
If one expands the amplitude first to ${\cal O}(\omega^2)$, the coefficients
still depend on $M_\pi$. Expanding those coefficients in a second step
in $M_\pi$ so that the overall order is ${\cal O}(M_\pi^2)$ for
$\omega = M_\pi$, one obtains the original LEG \cite{David}. The
mathematical origin of the problem is an illicit interchange of limits~:
expanding a function $f(\omega,M_\pi)$ in the manner just described
and setting $\omega = M_\pi$ at the end will in general not lead
to the same result as an expansion of $f(M_\pi,M_\pi)$ to the
same order in $M_\pi$.

Although it is shown in Ref.~\cite{David} that one can recover the
correct LET by a resummation of the series to all orders in $\omega$,
there is in general no guarantee that off--shell manipulations
produce the correct result. A simple, but
instructive example is to consider the elastic $\pi\pi$
scattering amplitude to lowest order, ${\cal O}(p^2)$, both in CHPT and
in the linear $\sigma$ model. Although the amplitudes agree on--shell,
they disagree in general off--shell. In fact, one can obtain very
different forms for the off--shell amplitude by redefining
the pion field. While one would normally not employ such
redefinitions in the linear model (seemingly destroying
renormalizability), any choice of pion field is equally acceptable
in CHPT which is based on an intrinsically non--renormalizable
quantum field theory.

Off--shell manipulations are dangerous and may lead to incorrect
results. The literature on applications of current algebra techniques
abounds with examples.

\subsection*{(d) Phenomenology}
Although the purpose of this comment is not to discuss the experimental
situation, it may be one of nature's follies that
experiments seem to favour the original LEG over the correct LET.
One plausible explanation for the seeming failure of the
LET is the very slow convergence of the expansion in $M_\pi$ \cite
{BKM1} \cite{vero}. CHPT produces a satisfactory
description of the total and differential cross sections
near threshold \cite{sac} \cite{beck} \cite{BKM1} \cite{vero}.
On the other hand, the extrapolation
to threshold involves sizable isospin violating corrections
that are not fully under control \cite{BKM2}. Both for the isospin
violating corrections and for the slow convergence of the expansion in
$M_\pi$, the LET for $\pi^0$ photoproduction at threshold does not seem
to be the ideal place to test the SM. It will however remain
an important theoretical check for any model of hadronic interactions
at low energies.

\vspace{1cm}

\section*{Acknowledgements}
Talks with the title of this comment were presented by G.E. at the
Workshop on Chiral Dynamics~: Theory and Experiment, MIT, Cambridge,
July 1994 and by U.-G.M. at the Gordon Research Conference on
Photo--Nuclear Reactions, Tilton, New Hampshire, August 1994. We are
grateful to many participants of those meetings encouraging us
to make this material available in written form. We are also
indebted to many friends and collaborators for their comments
and for their work described here, especially V\'{e}ronique
Bernard, J\"urg Gasser, Norbert Kaiser, Joachim Kambor and Heiri
Leutwyler. Finally, we thank J\"urg Gasser for a critical
reading of the manuscript.

\vfill
\section*{Figure Captions}
\begin{description}
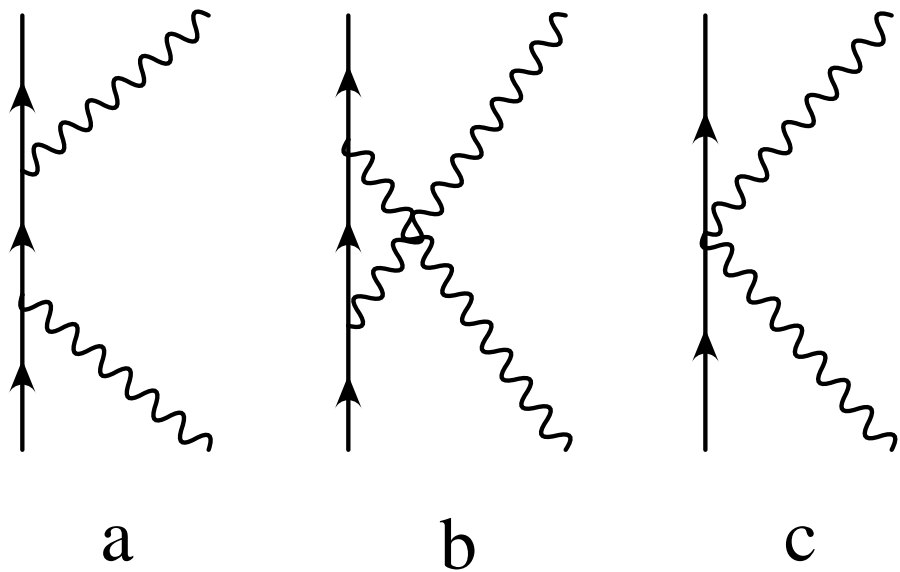
\item[Fig. 1:] Tree diagrams of ${\cal O}(p)$ (a,b) and ${\cal O}(p^2)$
(c) for Compton scattering in HBCHPT. Full (wavy) lines stand for
nucleons (photons).
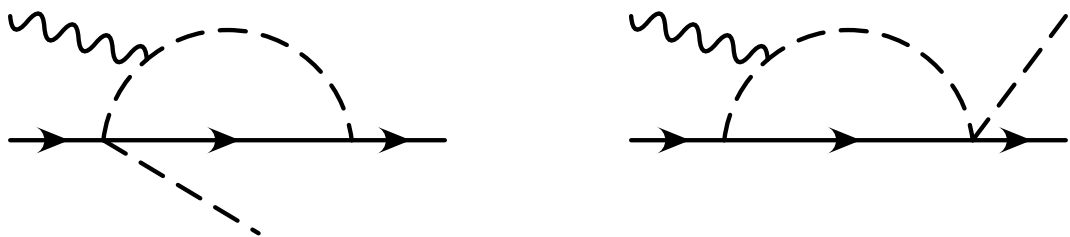
\item[Fig. 2:] One--loop triangle diagrams contributing to the threshold
amplitude $E_{0+}$ for $\pi^0$ photoproduction at ${\cal O}(M_\pi^2)$.
Pions are denoted by broken lines.
\end{description}

\end{document}